\documentclass[twocolumn,aps,prb,showpacs,amsmath,superscriptaddress,longbibliography,notitlepage]{revtex4-2}
\usepackage{amssymb,mathtools}
\usepackage[english]{babel}
\usepackage{mathrsfs}
\usepackage{graphicx}
\usepackage{float}
\usepackage{pbox}
\usepackage[caption=false]{subfig}
\usepackage{dcolumn}
\usepackage{appendix}
\usepackage{tabularx}
\usepackage{txfonts}
\usepackage[normalem]{ulem}
\usepackage{verbatim}
\usepackage{xcolor}
\usepackage{bm}
\usepackage{natbib}
\usepackage{multirow}
\usepackage{soul}

\begin{document}
\title{Engineering  Higher-Order Dirac and Weyl Semimetallic phase in 3D Topolectrical Circuits}
\author{S. M. Rafi-Ul-Islam}
\email{rafiul.islam@u.nus.edu}
\affiliation{Department of Electrical and Computer Engineering, National University of Singapore, Singapore 117583, Republic of Singapore}
\author{Zhuo Bin Siu}
\email{elesiuz@nus.edu.sg}
\affiliation{Department of Electrical and Computer Engineering, National University of Singapore, Singapore 117583, Republic of Singapore}
\author{Haydar Sahin}
\email{sahinhaydar@u.nus.edu}
\affiliation{Department of Electrical and Computer Engineering, National University of Singapore, Singapore 117583, Republic of Singapore}
\affiliation{Institute of High Performance Computing, A*STAR, Singapore 138632, Republic of Singapore}
\author{Mansoor B.A. Jalil}
\email{elembaj@nus.edu.sg}
\affiliation{Department of Electrical and Computer Engineering, National University of Singapore, Singapore 117583, Republic of Singapore}

\begin{abstract}
We propose a 3D topolectrical (TE) network that can be tuned to realize various higher-order topological gapless and chiral phases. We first study a higher-order Dirac semimetal phase that exhibits a hinge-like Fermi arc linking the Dirac points. This circuit can be extended to host highly tunable first- and second-order Weyl semimetal phases by introducing a non-reciprocal resistive coupling in the $x-y$ plane that breaks time reversal symmetry. The first- and second-order Weyl points are connected by zero-admittance surface and hinge states, respectively. We also study the emergence of first- and second-order chiral modes induced by resistive couplings between similar nodes in the $z$-direction. These modes respectively occur in the midgap of the surface and hinge admittance bands in our circuit model without the need for any external magnetic field.    
\end{abstract}
\maketitle

\subsection{Introduction}
Topological materials can be classified as either gapped or gapless based on their energy band spectra in momentum space \cite{vergniory2019complete,xu2023photoelectric,PhysRevResearch.4.013243,yan2017topological,Xu_2023,10.3389/fphy.2022.1021192,islam2014thermal}. The former hosts many exotic phenomena ranging from topological insulators \cite{hasan2010colloquium,rafi2024twisted,qi2011topological}, integer quantum Hall insulators \cite{shimshoni1997quantized,sun2020spin,he2017chiral,sun2019field}, and topological superconductors \cite{bernevig2013topological,sato2017topological} to higher-order topological insulators  \cite{schindler2018higher,ezawa2018higher}. These gapped topological phases are characterized by topological invariants such as the Chern number \cite{hatsugai1993chern}, Berry phase \cite{fujita2011gauge,vanderbilt2018berry}, and $Z_2$ invariant \cite{kane2013topological}. In contrast, gapless topological systems are characterized by the nature of their band degeneracy points where two or more bands touch one other in momentum space. These band degeneracy nodes are classified as either Dirac points (DPs) \cite{tarruell2012creating} or Weyl points (WPs) \cite{lu2013weyl,rafi2023conductance} depending on their symmetries. Dirac points emerge only when both time-reversal and inversion symmetries are present in a system. In contrast, Weyl points appear in the band dispersion if either or both symmetries are broken. Both types of band touching points appear and annihilate pairwise. Two important classes of topological systems that host WPs and DPs are Weyl semimetals (WSMs) \cite{yan2017topological,rafi2020strain,lv2015experimental,rafi2020realization,vazifeh2013electromagnetic} and Dirac semimetals \cite{young2015dirac,armitage2018weyl}, respectively.

Recently, a new class of three-dimensional (3D) topological phases named higher-order topological insulators (HOTI), which go beyond the usual bulk-boundary correspondence, has been discovered \cite{schindler2018higher}.  In general, a $d$-dimensional $n$th-order topological insulator can host topologically protected $(d-n)$-dimensional gapless boundary states \cite{hofmann2020reciprocal,longhi2019probing,ezawa2019braiding}. Higher-order topological insulators are insulating in the bulk or surfaces and become metallic only when edges or hinges are introduced, respectively. They present intriguing multidimensional topological phenomena ranging from corner states to hinge states \cite{schindler2018higher,ezawa2018higher,kempkes2019robust,xie2018second}. Interestingly, such unconventional non-trivial boundary modes are robust against system disorders and are protected by certain crystalline symmetries (e.g., reflection and mirror symmetries).   Meanwhile, Weyl and Dirac semimetals have isolated band-touching points and exhibit unconventional properties such as the chiral anomaly and, in particular, Fermi arcs \cite{zyuzin2012topological,siu2017influence,pikulin2016chiral,xu2015discovery}.
However, owing to the difficulties involved in finding suitable materials and the complexity in tuning model parameters, only a few experimental realizations of higher-order Weyl semimetals (HOWSMs) \cite{ghorashi2020higher,wang2020higher} and Dirac semimetals (HODSMs) \cite{wang2020higher} have been reported so far in acoustic crystal systems\cite{wei2021higher,wang2022higher,borisenko2014experimental}. Interestingly, other higher-order nontrivial topological phases with unconventional topological bandstructures have been proposed in a multitude of platforms, e.g., in photonic \cite{xie2019visualization,liu2019second}, mechanical \cite{susstrunk2015observation}, and acoustic \cite{yang2015topological} systems, and in ultra-cold atomic gases in optical lattices \cite{lewenstein2007ultracold,sun2012topological}, polaritons \cite{zhang2020nonlinear}, \cite{chen2020plasmon}, micro-cavities \cite{edvardsson2019non}, optical waveguides and fibres \cite{xie2019visualization,luo2019higher}, non-hermitian systems \cite{PhysRevB.106.245128,PhysRevResearch.4.043108,rafi2022critical} and others \cite{nash2015topological}.  Each of these platforms comes with experimental complexities and drawbacks, which makes them vulnerable to perturbations and non-uniformities. 

In the search for alternative platforms to serve as experimental testbeds for investigating topological states, lattice arrays with lossless electrical components such as inductors and capacitors known as topolectrical (TE) circuits have emerged as a frontrunner \cite{olekhno2020topological,zhang2023anomalous,sahin2022impedance,rafi2020realization,helbig2020generalized,rafi2020topoelectrical,helbig2019band,rafi2020anti,imhof2018topolectrical,rafi2023valley,lee2018topolectrical,rafi2021topological,hofmann2019chiral,rafi2021non,bao2019topoelectrical,rafi2021unconventional,zhang2022anomalous,rafi2022system,rafi2022interfacial} as they offer better ability for tuning and modulating the system parameters. Because TE circuits are not constrained by physical dimensionality but rely solely on the mutual connectivities between the voltage nodes, HOTIs and higher-order gapless systems \cite{ghorashi2020higher,wang2020higher} (i.e, HODSMs and HOWSMs) can be readily implemented using conventional electrical components. The gapless points in HODSMs and HOWSMs are protected by crystal symmetries and the WPs are connected by higher-order hinge-like Fermi arc states rather than conventional surface arc states \cite{ghorashi2020higher}. This suggests that highly robust hinge states can be achieved on the TE platform.

In this paper, we propose TE circuit networks that host HOWSM and HODSM non-trivial states that can be switched on and tuned solely by the choice of circuit parameters.  We first construct a prototypical 2D TE circuit model which exhibit the HOTI phase. To realize the gapless HOWSM and HODSM phases, we then stack copies of the 2D circuit lying on the $x$-$y$ plane on the top of one another along the $z$-direction and couple the adjacent layers diagonally via a common stacking capacitor $C_z$. The stacking capacitor has the effect of modifying the intra- and intercell hopping in the effective 2D Laplacian as well as introducing an additional $k_z$ dependence. Because the 3D circuit still obeys time-reversal and inversion symmetry, the circuit hosts pairs of Dirac points with higher-order topology. These symmetries can be broken by introducing a non-reciprocal resistive coupling that connects the nodes within a unit-cell diagonally on the $x$-$y$ plane. The symmetry breaking results in the emergence of first- and second-order Weyl points connected by a zero-admittance flat band, similar to surface and hinge Fermi arcs \cite{wieder2020strong,wang2020vortex}, respectively. A tilting capacitor $C_t$ that connects the same types of nodes along the $z$ axis can be further introduced to give rise to a tilted admittance dispersion while retaining the higher-order topology. A signature of these higher-order topologies is the localization of the squared amplitude of the nodal voltages, which is the TE equivalent of the quantum mechanical particle density, along the hinges of a 3D system having a nanowire geometry, i.e., with open boundary conditions along two dimensions. Finally, the chiral symmetry of the circuit lattice can be broken by introducing loss (positive resistance) and gain (negative resistance) terms between the same type of nodes in adjacent layers. This will result the emergence of midgap chiral surface and hinge states in the midgap of the admittance spectra. These chiral modes are resilient against system perturbations and disorders.  Therefore, both first- and second-order chiral states can be induced in the proposed TE circuit without any external magnetic field. These novel higher-order topologically non-trivial chiral states may find many applications in fault tolerant quantum computing \cite{you2019higher}, robust signal multiplexing \cite{ni2020robust}, and dissipationless interconnects \cite{mei2019robust}.

\section{Results}
\subsection{Topolectrical Model}
\begin{figure*}[ht!]
  \centering
    \includegraphics[width=0.75\textwidth]{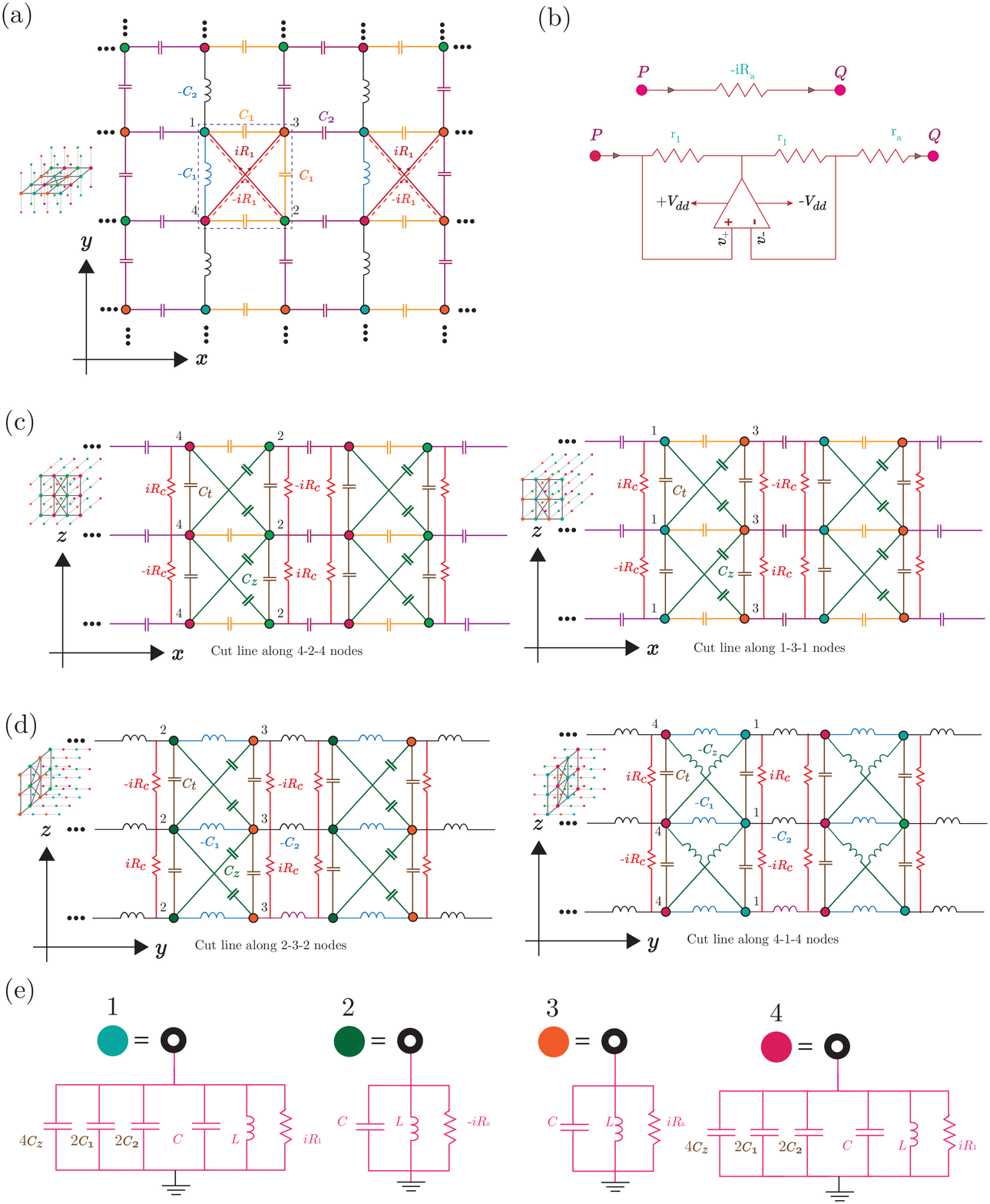}
  \caption{Schematic of a TE circuit hosting higher-order semimetal states. (a) Cross-section of the TE lattice model on the $x$-$y$ plane. The blue, green, orange, and magenta circles represent the $1$, $2$, $3$ and $4$ sublattice sites or nodes, respectively. The dashed rectangle delineates a unit cell. The cartoon at the left of the lattice model schematically illustrates the crystal plane depicted. The intracell and intercell couplings along the $x$ and $y$ axes are given by capacitors $C_1$ and $C_2$, respectively. Note that there is an additional $\pi$-phase shift in the coupling linking the $4$-$1$-$4$ nodes along the $y$-axis compared to that linking the $2$-$3$-$2$ nodes. The negative capacitance represents a frequency-dependent inductance (i.e., $-C_i=(\omega^2 L_i)^{-1}$). Non-reciprocal resistive couplings $\pm iR_d$ link diagonal nodes within the unit cell, giving rise to the breaking of time-reversal symmetry which is a requirement in realizing the Weyl-semimetal phase. (b) Negative impedance converter for providing an extra phase shift of $\pi$ (change of sign) to the impedance and therefore converting a lossy resistive term $R_a$ to a gain term $-R_a$. (c) Cross-section of the TE lattice model on the $x$-$z$ plane with the cut line at two different positions. The circuit is extended along the vertical $z$-direction by stacking layers of the circuit lattice on the $x$-$y$ plane described in (a) using capacitors, inductors and resistors. Nodes 2 and 4, which are diagonal to each other, are connected by a common capacitor $C_z$. (d) Cross-section of the TE lattice model on the $y$-$z$ plane with the cut line at two different positions. The diagonal nodes of the cut lines along the $4-1-4$ and $2-3-2$ nodes are connected by an inductor $-C_z$ and capacitor $C_z$, respectively. Note that, for (c) and (d), the same type of nodes are connected along $z$-axis by a resistive element $R_c$ with alternating signs and a tilting capacitance $C_t$. (e) Grounding mechanism of the TE circuit for all four types of nodes. The common grounding capacitor $C$ serves the role of the eigenenergy while the common inductor $L$ is added to make the momentum-independent diagonal elements in the Laplacian matrix zero, which is analogous to setting the onsite energy to zero for a condensed matter tight-binding Hamiltonian. }
  \label{gFig1}
\end{figure*}

To realize higher-order topological gapless states, we consider a three-dimensional TE circuit consisting of inductors, capacitors, resistors, and operational amplifiers, as shown in Fig. \ref{gFig1}. The TE circuit has a unit cell (indicated by the dashed box in Fig. \ref{gFig1}a ) consisting of four sublattice nodes denoted as $\mathrm{1}$, $\mathrm{2}$, $\mathrm{3}$, and $\mathrm{4}$. The intracell and intercell couplings on the $x-y$ plane are given by the capacitances of $C_1$ and $C_2$, respectively. The coupling strength linking nodes $4$-$1$-$4$ in the $y$-direction has a negative sign, denoting the inductive nature of the coupling (i.e., $-|C_i|=(\omega^2 L_i)^{-1}$, where $\omega$ is the frequency of the driving alternating current in the circuit). Additionally, the resistive couplings between the diagonal nodes within a unit cell are non-reciprocal and direction-dependent, and are given by $iR_d$ ($-i R_d$) for the solid (dashed) lines (Fig. \ref{gFig1}a) where the resistive couplings in capital letter (e.g., $R_j$) are related to the physical resistance $r_j$ through $R_j=1/(i \omega r_j)$. Note that the positive and negative resistive elements in a TE circuit correspond to loss and gain terms in quantum mechanics. The $\pi$-phase shift or change of sign in the resistive coupling can be achieved by using the impedance converter set-up shown in Fig. \ref{gFig1}b. The combination of two identical resistors $r_1$ and an ideal operational amplifier with supply voltages $V_{dd}^{+}$ and $V_{dd}^{-}$ effectively changes the resistance between nodes P and Q from $r_a$ to $-r_a$, thus behaving as a negative resistance converter with a resistive coupling of $-iR_a$ (see Appendix A for more details).  These non-reciprocal loss ($i R_d$) and gain ($-i R_d$) terms would be crucial in breaking the time reversal symmetry of the circuit to allow the system to host higher-order Weyl points, as will be discussed later.
 
To induce the higher-order gapless states with richer topological properties compared to their first-order counterparts, it is necessary to extend the TE circuit vertically in the $z$-direction by stacking the two-dimensional $x$-$y$ layers of the circuit shown in Fig. \ref{gFig1}a. The nodes in the unit cell are coupled diagonally in the $x$-$z$ plane to the adjacent layer by a common capacitor $C_z$ (see Fig. \ref{gFig1}c). The diagonal couplings within a unit cell on the $y-z$ plane, which are of strength $-C_z$, are provided by an inductor (see Fig. \ref{gFig1}d). These inter-layer couplings of  $\pm C_z$ effectively modify the intracell couplings of the original admittance matrix of the two-dimensional $x$-$y$ circuit layer from $C_1 \to C_1+2C_z \cos k_z$ if we regard the vertical wavevector $k_z$ as a model parameter. This modification translates the circuit Laplacian into a mathematically equivalent 2D SSH model, as will be discussed later. Additionally, the same types of nodes are connected to the adjacent vertical layers by a common tilting capacitance $C_t$. Aside from the inter-layer capacitive/inductive couplings, there are also resistive ones. Nodes 1 and 4 (nodes 2 and 3) are connected to the corresponding nodes in the upper adjacent layer by a positive (negative) resistive coupling $i R_c$ ($-i R_c$), respectively. These resistive couplings adopt the opposite sign when coupling to the lower adjacent layer (see Figs. \ref{gFig1}c and \ref{gFig1}d). Finally, all the electrical nodes are connected to ground by a common inductor $L$ and capacitor $C$ (see Fig. \ref{gFig1}e). The capacitance $C$ plays the role as the eigenenergy analogous to the Schr\"{o}dinger equation \cite{rafi2020anti,rafi2020realization,rafi2020topoelectrical}, while the common inductance $L$ allows tuning of the resonant frequency of the circuit.  The other capacitors and resistors connecting each node to the ground (see Fig. \ref{gFig1}e) ensure that the diagonal elements in the Laplacian matrix have the desired form as presented later in Eq. \eqref{WSMham}\cite{rafi2020realization}.

The dynamics of the TE circuit depicted in Fig. \ref{gFig1} can be described in reciprocal space by the four-band circuit Laplacian 
\begin{widetext}
\begin{eqnarray}
	&& Y(\omega,\boldsymbol{k}) =  i \omega\left((C_1+2C_z \cos k_z+C_2 \cos k_x)\sigma_x \otimes \sigma_0 +C_2 \sin k_x \sigma_y \otimes \sigma_z + (C_1+2C_z \cos k_z+C_2 \cos k_y)\sigma_y \otimes \sigma_y \right.) \nonumber \\ && + \left.( C_2 \sin k_y \sigma_y \otimes \sigma_x + i R_1 \gamma_1 + 2 R_c \sin k_z \sigma_z \otimes \sigma_z + 2 C_t \cos k_z \sigma_0 \otimes \sigma_0 -(2(C_1+C_2+2C_z)-(\omega^2 L)^{-1})\sigma_0 \otimes \sigma_0  \right) ,
 	\label{WSMham}   	
\end{eqnarray}
\end{widetext}
where $\sigma_i$ denotes the $i$th Pauli matrices in the sublattice space,  $\sigma_0$ is the $2\times 2$ identity matrix, $\gamma_1=(\sigma_y \otimes \sigma_x)(\sigma_y \otimes \sigma_z)$, $C_i (-C_i)$ is the coupling capacitance (inductance), and $R_1$ and $R_c$ are the resistive coupling strengths that break time-reversal and chiral symmetry, respectively. We set the resonant frequency in such a way that the $\mathbf{k}$-independent coefficients of the identity matrix vanish (i.e, $\omega_r=1/\sqrt{2L(C_1+C_2+2C_z)}$). This is equivalent to setting the onsite energy to zero at each lattice site in a tight-binding Hamiltonian \cite{savin2003heat}.

\subsection{Higher Order Dirac Semimetals: $R_1=0$, $R_c=0$,  $C_t=0$}
\begin{figure*}[ht!]
\centering
\includegraphics[width=0.85\textwidth]{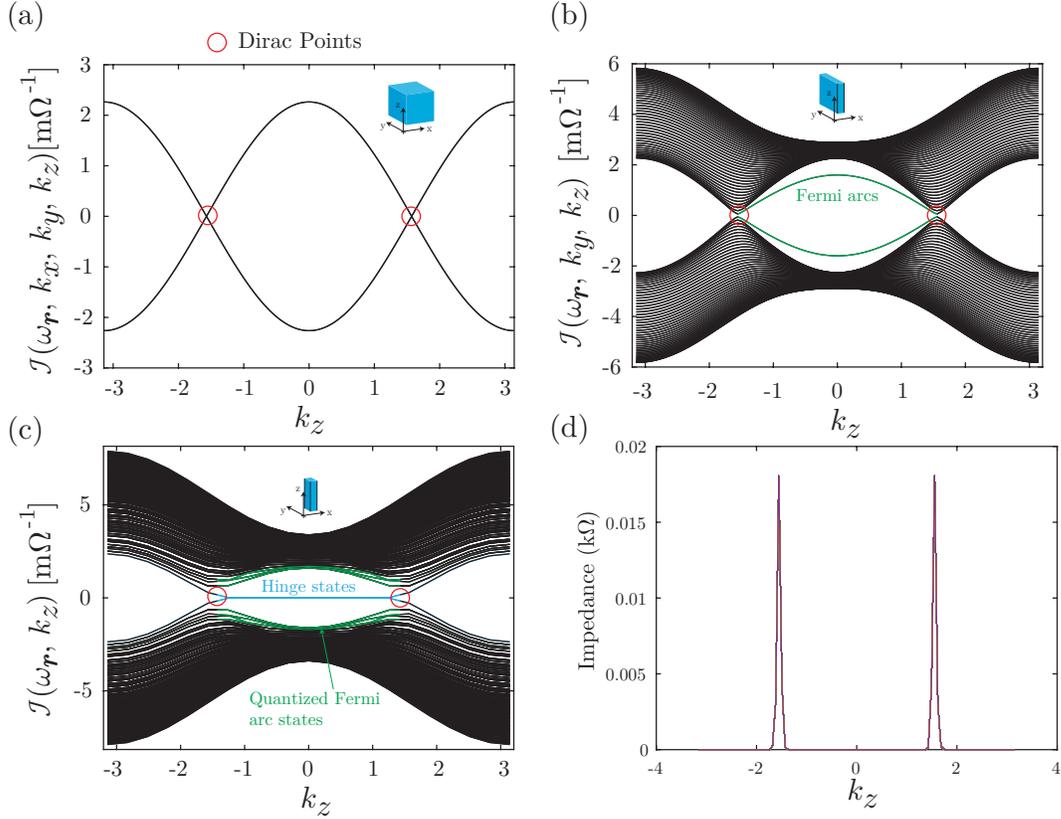}
\caption{ (a) Admittance band structure for the infinite (bulk) TE circuit model described in Eq. 1 corresponding to $k_x=0$, $k_y=0$, $R_d=0$, $R_c=0$ and $C_t=0$. Clearly, there exists a pair of Dirac points on the $k_z$ axis (indicated by red circles). (b) Admittance spectrum of the TE circuit in a nanoplate geometry consisting of 20 unit cells in the finite $x$-direction and PBC in the infinite $y$ and $z$ directions with $k_y=0$. Note the two Fermi arc edge states connecting the Dirac points. (c) The admittance dispersion of the TE circuit in the nanowire geometry with PBC along the $z$ direction and OBC in both the $x$ and $y$ directions. Note the flat band hinge state connecting two Dirac points. (d) Impedance spectrum as a function of $k_z$ for the case described in (b). The impedance is measured between the two terminal nodes. The common parameters are $C_1=-2$ $\mathrm{mF}$, $C_2= 2$ $\mathrm{mF}$, $C_z= 0.8$ $\mathrm{mF}$, $R_c=0$, and $C_t=0$.}
\label{gFig2}
\end{figure*}

In the absence of the resistive couplings and the tilting capacitance $C_t$ and at resonant frequency, the TE circuit satisfies chiral, inversion, reflection (across all three of the $x$, $y$ and $z$ axes) and mirror rotation symmetries. Mathematically, the circuit Laplacian in Eq. \ref{WSMham} satisfies the following
\begin{equation}
\begin{aligned}
\mathcal{C} Y (\omega_r,k_x,k_y,k_z)\mathcal{C}^{-1} &= - Y (\omega_r,k_x,k_y,k_z)\\
\mathcal{I} Y (\omega_r,k_x,k_y,k_z)\mathcal{I}^{-1} &=  Y (\omega_r,-k_x,-k_y,-k_z)\\
\mathcal{T} Y (\omega_r,k_x,k_y,k_z)/(i\omega) \mathcal{T}^{-1} &=  Y^{\dagger}(\omega_r,-k_x,-k_y,-k_z) /(i\omega)\\
\mathcal{M}_x Y (\omega_r,k_x,k_y,k_z){\mathcal{M}_x}^{\dagger} &=  Y (\omega_r,-k_x,k_y,k_z)\\
  \mathcal{M}_y Y (\omega_r,k_x,k_y,k_z){\mathcal{M}_y}^{\dagger} &=  Y (\omega_r,k_x,-k_y,k_z)\\
  \mathcal{M}_z Y (\omega_r,k_x,k_y,k_z){\mathcal{M}_z}^{\dagger} &=  Y (\omega_r,k_x,k_y,-k_z)\\
   \mathcal{M}_{xy} Y (\omega_r,k_x,k_y,k_z){\mathcal{M}_{xy}}^{\dagger} &=  Y (\omega_r,k_y,k_x,k_z) ,
\end{aligned}
\label{eq2}
\end{equation} where $\mathcal{C}$, $\mathcal{I}$, $\mathcal{T}$, $\mathcal{M}_i$ and $\mathcal{M}_{xy}$ are the chiral, inversion, time reversal, reflection about the $i$th axis with $i\in\{x,y,z\}$, and the $xy$ mirror rotation (i.e., about the $\hat{x}+\hat{y}$ direction) operators, respectively. In terms of the Pauli matrices, these operators are explicitly given by $\mathcal{C}=\sigma_z \otimes \sigma_0$, $\mathcal{I}=\sigma_0 \otimes \sigma_y$, $\mathcal{T}=\sigma_0 \otimes \sigma_y \kappa$ with $\kappa$ denoting complex conjugation, $\mathcal{M}_x=\sigma_x \otimes \sigma_z$, $\mathcal{M}_y=\sigma_x \otimes \sigma_x$, $\mathcal{M}_z=\sigma_0 \otimes \sigma_0$, and $\mathcal{M}_{xy}=\frac{1}{2}(\sigma_{+} \otimes \sigma_0 + \sigma_{-}\otimes i \sigma_y )\cdot(\sigma_x \otimes \sigma_x))$, with $\sigma_{\pm}=\sigma_x \pm i\sigma_y$.
In the absence of the resistive couplings and tilting capacitance $C_t$, the admittance eigenvalues of Eq. \ref{WSMham} for the bulk periodic system take the form of 
\begin{eqnarray}
&&\mathcal{J}(\omega,k_x, k_y,k_z)\nonumber \\
&=& \pm\omega \sqrt{ \sum_{i=x,y} \left( (C_1+2C_z \cos k_z +C_2 \cos k_i)^2 +(C_2 \sin k_i)^2 \right)}.
\label{eq3}  
\end{eqnarray}
The admittance band gap closes at $(k_x,k_y,k_z)=(0,0,{\kappa_z}^{D1})$ and $(k_x,k_y,k_z)=(0,0,{\kappa_z}^{D2})$, where ${\kappa_z}^{D1}=\pm \cos^{-1} \left(-\frac{C_1+C_2}{2C_z}\right)$, and ${\kappa_z}^{D2}=\pm \cos^{-1} \left(-\frac{C_2-C_1}{2C_z}\right)$. 
We consider the first case where the bands touch at $\kappa_z^{D1}$ (a similar behaviour occurs for the other case where band gap closing occurs at $\kappa_z^{D2}$). If $|C_1+C_2| < |2C_z|$, the inversion and time-reversal symmetries guarantee the existence of a pair of Dirac points in $k$-space for the bulk admittance band structure (see Fig. \ref{gFig2}a). The imposition of a finite width, i.e., open boundary conditions (OBC), on this system along the $x$-direction while retaining periodic boundary conditions (PBC) along the $y$ and $z$-directions results in a nanoplate geometry. Surface edge states localized near the $x$ boundaries of the nanoplate now emerge and connect the pair of Dirac points (see Fig. \ref{gFig2}b). These midgap edge modes are Fermi arc states that connect two gapless points with opposite topological charges in conventional (first-order) Weyl semimetals \cite{rafi2020realization,lv2015experimental}. Next, we now further impose a finite width on the $y$ direction, so that a nanowire geometry with OBC in the $x$ and $y$-directions and PBC in the $z$-direction is formed. The Fermi arc states in the nanoplate geometry now become quantized into sub-bands because of the geometrical confinement along the $y$ direction, and no longer cross the gap (Fig. \ref{gFig2}c). A nearly-flat state linking the two Dirac points (DPs) appears in the admittance dispersion along the $k_z$ axis (see Fig. \ref{gFig2}c). This state is a higher-order state because it emerges only when the system is confined along two directions. We shall also explicitly show later that it has a higher-order topology by calculating its topological number and showing its localization along the hinges of the nanowire. 
Because the Laplacian obeys both chiral $\mathcal{C}$ and mirror rotational symmetries $\mathcal{M}_{xy}$ (see Eq. \ref{eq2}), the Laplacian can be transformed via a unitary transform into a block diagonal form along the $k_x=k_y=k$ high symmetry line \cite{schindler2018higher,kempkes2019robust}. The transformed Laplacian is given by
\begin{eqnarray}
\chi ^{-1} Y (\omega_r,k,k_z) \chi = \begin{bmatrix}
Y_1 (\omega_r,k,k_z) & 0\\
0 & Y_2 (\omega_r,k,k_z)
\end{bmatrix}
,\label{eq4}
\end{eqnarray}
where $\chi$ is a unitary transformation. The diagonal elements of the transformed block matrix \cite{liu2019second} can be expressed as 
\begin{eqnarray}
\begin{aligned}
Y_{1,2}(\omega_r,k,k_z)=&(C_1+2C_z \cos k_z+C_2 \cos k) \sigma_x  \\ & + \eta_{1,2} C_2 \sin k \sigma_y,
\label{eq5}
\end{aligned}
\end{eqnarray}
where, $\eta_{1,2}=1$ $(-1)$ for $Y_1$ $(Y_2)$ respectively. (The fact that the matrix in Eq. \eqref{eq4} is related to $Y(\omega_r,k,k_z)$ via a unitary transform can be verified by noting that they share the same set of eigenvalues.) Interestingly, Eq. \ref{eq5} represents a modified SSH model with a $k_z$-dependent intracell coupling. Therefore, the resultant $(4\times 4)$ admittance matrix in Eq. \ref{eq4} can be regarded as two decoupled SSH blocks. If we consider $k_z$ as a model parameter, the winding number ($\mathcal{W}$) of the TE circuit, which serves as its topological index, can be obtained (see Appendix B for details) as
\begin{equation}
\mathcal{W}=
\begin{cases}
  2, & \text{if }
       \!\begin{aligned}[t]
       \bigg|\frac{C_1+2C_z \cos k_z}{C_2}\bigg| \leq 1 \\
            \end{aligned}
\\
  0, & \text{otherwise}
\end{cases}.
\label{eq6}
\end{equation}
The above results thus show that the parameter regime with flat hinge states connecting the two DPs in Fig. \ref{gFig2}c carries a topological index of two, confirming its character as a higher-order hinge Fermi arc. We can thus classify the circuit as a higher-order Dirac semimetal (HODSM).

In addition to the characterization of the HODSM state based on its admittance spectrum and topological index, we can also obtain a signature of the higher-order topological state from the impedance spectrum of the TE circuit. In general, the impedance between any two arbitrary nodes $m$ and $n$ in the circuit can be measured by connecting an external current source providing a fixed current $I_{mn}$ to the two nodes and measuring the resulting voltages at the two nodes $V_m$ and $V_n$. The impedance is then given by
\begin{equation}
Z_{mn}=\frac{V_m-V_n}{I_{mn}}=\sum_i \frac{|\psi_{i,m}-\psi_{i,n}|^2}{\lambda_i},
\label{eq7}
\end{equation}
where $\psi_{i,a}$ and $\lambda_i$  are the voltage at node $a$ and the eigenvalue of the $i^{\mathrm{th}}$ eigenmode of the (finite-width) Laplacian, respectively. Fig. \ref{gFig2}d shows the impedance variation of the  HODSM for OBC along the $x$ direction and PBC along the $y$ and $z$ directions. The impedance is plotted as a function of $k_z$ for the measurement between the two terminal points. The peak positions of the high-impedance states exactly match that of the higher-order DPs in Fig. \ref{gFig2}a. In the contrary, the high impedance peaks occur at the phenomenal points \cite{rafi2022interfacial,zhang2022universal,rafi2021topological} in the second-order gapless points (see Appendix C for details). This suggests the possible electrical characterization of higher-order DPs in the TE system via impedance measurements on the circuit. In addition, we thoroughly discussed the distinction between first order and second order gapless states using impedance spectra in Appendix C.

\subsection{Chiral Hinge States in HODSM: $R_1=0$, $R_c\neq 0$, $C_t=0$}\begin{figure*}[ht!]
\centering
\includegraphics[width=0.85\textwidth]{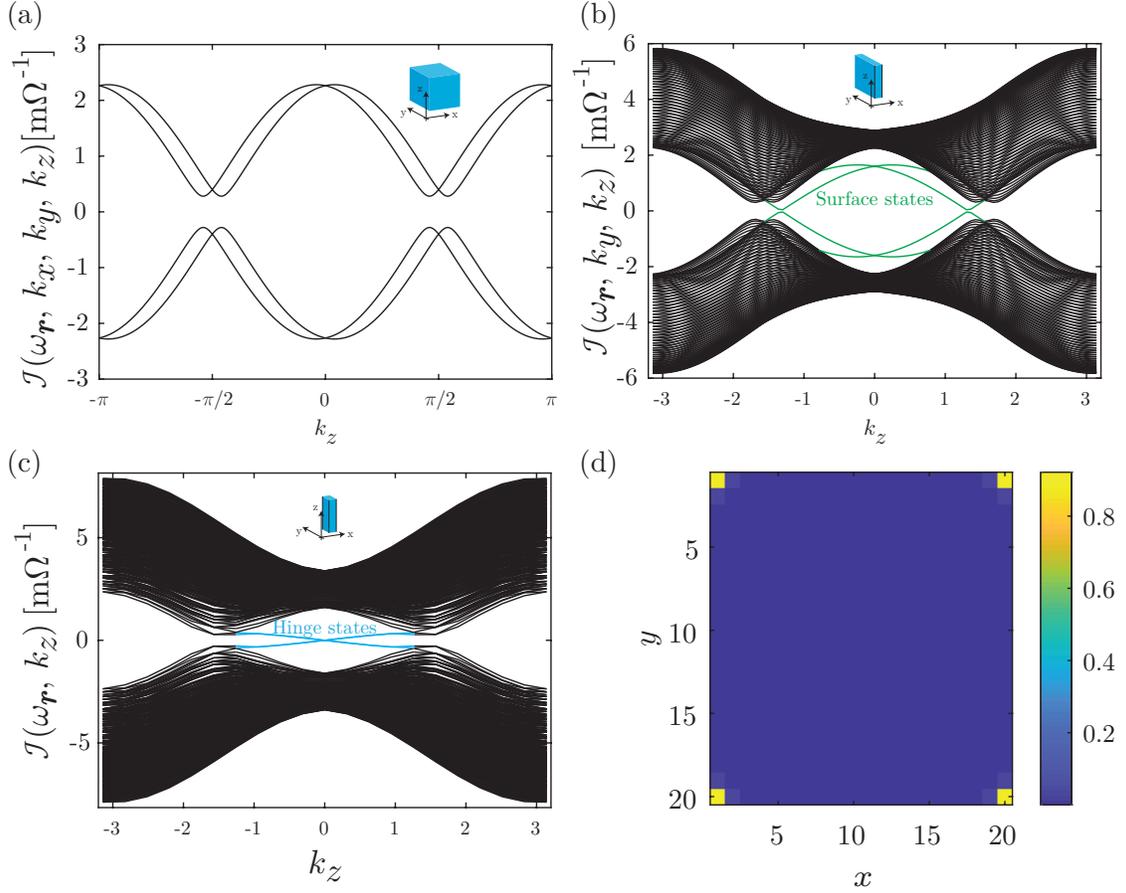}
\caption{ Bulk admittance band dispersion for the TE model described in Eq. \eqref{WSMham} with finite $R_c$. The other parameters are $k_x=0$, $k_y=0$, $R_d=0$, and $C_t=0$. A non-zero value of $R_c$ will induce a finite gap in the spectrum, and there will thus be no Dirac points. (b) Admittance spectrum of a circuit consisting of 20 unit cells along the $x$ axis with OBC, and PBC along the $y$ and $z$-directions with $k_y=0$. Note that the surface states become gapped and do not form Dirac points. (c) Chiral topological hinge states appear when OBCs are imposed in both the $x$ and $y$-directions. (d) Squared nodal voltage amplitude distribution for a nanoplate geometry with 20 unit cells along both the $x$ and $y$-directions, and with $k_z=0$. Only the nodes at the corner exhibit significant voltages, indicating the presence of topological conducting hinge states. The common parameters are $C_1=-2$ $\mathrm{mF}$, $C_2= 2$ $\mathrm{mF}$, $C_z= 0.8$ $\mathrm{mF}$,  $R_c=0.8$ $\mathrm{mF}$ and $C_t=0$.}
\label{gFig3}
\end{figure*} 
 In the previous section, we discussed the evolution of HODSM phases in a $\mathrm{LC}$ circuit lattice. In this subsection, we will study the effect of resistive coupling along the $z$-direction  (in the form of a finite $R_c$ in Eq. \ref{WSMham}) on the higher-order topology of the circuit model. A non-zero $R_c$ in Eq. \ref{WSMham} corresponds to alternating $i R_c$ and $-iR_c$ couplings between adjacent layers of the circuit along the $z$ direction, as shown in Fig. \ref{gFig1}b and Fig. \ref{gFig1}c. A finite $R_c$ breaks the reflection symmetry in the $z$-direction and that across the $x=y$ plane ($\mathcal{M}_{xy}$), as well as the chiral and time-reversal symmetries. However, it preserves inversion symmetry, as well as the reflection symmetries in the $x$ and $y$-directions.  
 
A finite $R_c$ breaks the degeneracy of the bulk admittance dispersion such that the two pairs of two-fold degenerate bands in the admittance spectrum split into four non-overlapping bands, as illustrated in Fig. \ref{gFig3}a. In addition, a finite band gap opens up. To further analyze the system, we expand the Laplacian in Eq. \ref{WSMham} in the vicinity of the DP at $(0,0,\kappa_1^{D1})$ for the case of $C_1=-C_2$, and obtain the low-energy Laplacian
\begin{widetext}
\begin{eqnarray}
Y_{Low}(\omega, q_x, q_y,q_z)=2C_z q_z \sigma_x \otimes \sigma_0 + C_2 q_x \sigma_y \otimes \sigma_z + 2C_z q_z \sigma_y \otimes \sigma_y  + C_2 q_y \sigma_y \otimes \sigma_x + 2 R_c \sigma_z \otimes \sigma_z.
\label{eq8}
\end{eqnarray}
\end{widetext}
From the above equation, we find that the admittance gap between the upper and lower bands in Fig. \ref{gFig3}a is given by $|4 R_c|$, which means that the Laplacian in Eq. \ref{eq8} describes a massive Dirac fermion. The edge states survive in the nanoplate geometry with OBCs in the $x$ axis and PBCs in the other two directions (see Fig. \ref{gFig3}b). Fig. \ref{gFig3}c shows the admittance spectra for the nanowire structure with OBCs in both the $x$ and $y$ directions and PBC in the $z$ direction. The imposition of the additional OBC along the $y$ direction on the nanoplate geometry quantizes the edge states, which no longer cross the band gap, and causes the emergence of a pair of chiral hinge states which cross at $k_z=0$ and closes the gap. To confirm the localization of the hinge states, we plot the squared amplitude of the node voltages for the nanowire geometry with 20 unit cells in both the $x$ and $y$ directions in Fig. \ref{gFig3}d. We find that at zero admittance, the nodes with significant voltages are confined to the corners or hinges. This is one signature of a higher-order topological insulator. The surface and higher order topological states can be characterized by different indices - the Chern number for the surface states and the quantized quadrupole moment index for the higher order states \cite{benalcazar2017electric,ghorashi2020higher}. For further information, please refer to Appendix D.

 \subsection{Tilted Chiral Hinge States : $R_1=0$, $R_c\neq 0$, $C_t\neq 0$}\begin{figure}[ht!]
\centering
\includegraphics[width=0.48\textwidth]{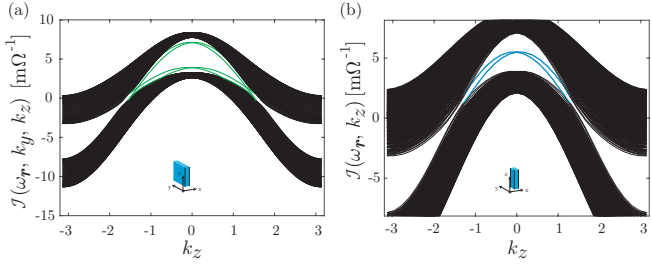}
\caption{ Type-2 chiral modes in TE model with a finite tilting capacitance $C_t$. (a) Surface admittance spectra for a TE circuit with OBC in the $x$ direction with 20 unit cells, and PBC in the $y$ and $z$ directions at $k_y=0$. (b) Admittance dispersion exhibiting Type-2 hinge modes in a TE circuit with OBC in both the $x$ and $y$ directions (with 20 unit cells along both these directions). Note that the two chiral hinge states propagate in a valley-dependent direction, i.e., they exhibit only positive (negative) group velocity in the $z$ direction at the $K$ ($K'$) valley. The common parameters are $C_1=-2$ $\mathrm{mF}$, $C_2= 2$ $\mathrm{mF}$, $C_z= 0.8$ $\mathrm{mF}$,  $R_c=0.2$ $\mathrm{mF}$, and $C_t= 2.75$ $\mathrm{mF}$.  } 
\label{gfig4}
\end{figure}	
We consider the effects of tilting on the admittance band dispersion. As can be seen from Eq. \ref{WSMham}, a finite value of $C_t$ leads to a tilt in the dispersion. In Fig. \ref{gfig4}, we plot the admittance spectra as a function of $k_z$ for non-zero $C_t$. The presence of tilt leads to a drastic modification of the edge and chiral hinge states. Interestingly, the edge states survive even when the whole spectra becomes overtilted when we consider a TE system with OBC in the $x$ direction and PBC in the $y$ and $z$ directions (whose dispersion is shown in Fig. \ref{gfig4}a). However, both edge states acquire the same sign of the admittance slope in the vicinity of each Dirac point. 

The chiral hinge modes that emerge when OBCs are imposed on both the $x$ and $y$ directions show some peculiar characteristics (see Fig. \ref{gfig4}b). The two chiral hinge states propagate in a direction that is valley-dependent.  At $k_z=0$, both chiral hinge modes have zero group velocity, but at finite values of $|k_z|$, they exhibit the same sign of the admittance slopes, as shown in Fig. \ref{gfig4}b. In other words, the hinge states in the $K$ ($K'$) valley propagate with positive (negative) group velocities in the $z$ direction . These overtilted higher-order edge and hinge states can be termed as Type-2 topological states \cite{rafi2020topoelectrical} and show a sharp contrast to the Type-1 surface and hinge states where both the $K$ and $K'$ valleys host states with both positive and negative group velocities.

\subsection{Higher-order Chiral Weyl Semimetals: $R_1\neq 0$, $R_c\neq 0$, $C_t = 0$}  
\begin{figure*}[ht!]
\centering
\includegraphics[width=0.85\textwidth]{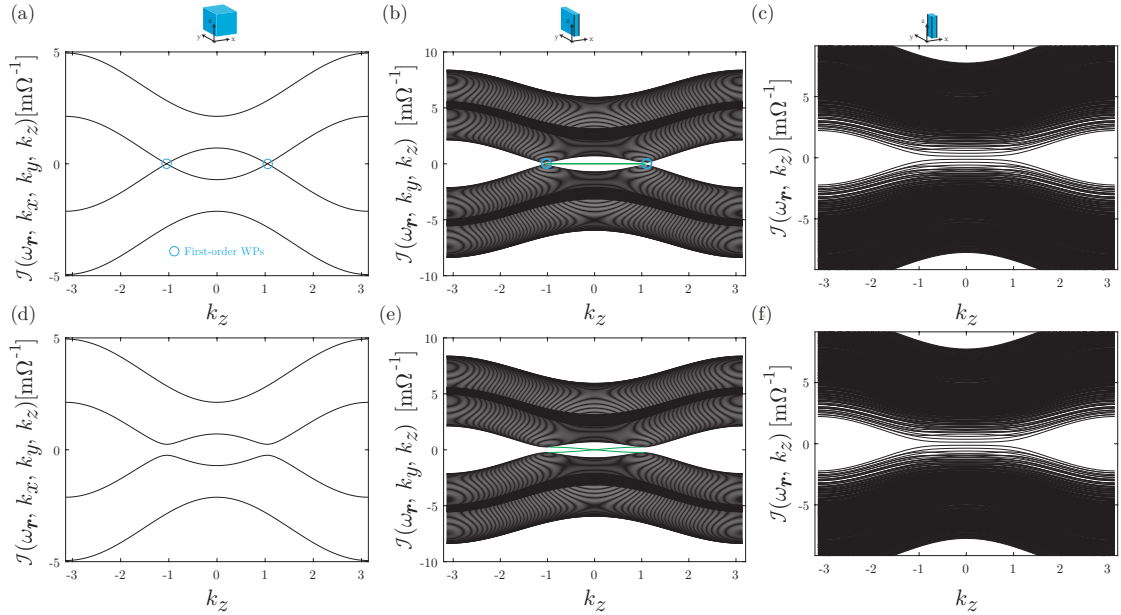}
\caption{ Evolution of the admittance band structure of the circuit model in Eq. \ref{WSMham} with non-zero resistive elements. (a) When $R_c=0$ and $k_x=k_y=0$, the low admittance bands cross each other at two first-order WPs in the Brillouin zone when $\eta=+1$ only. (b) The admittance spectra under open boundary conditions (OBC) in $x$ and periodic boundary conditions (PBC) in $y$ and $z$ directions with $k_y=0$. The WPs are connected by surface states with exactly admittance states, which close the gap on either side of $k_z=0$. (c) The admittance dispersion under OBC in $x$ and $y$ directions and PBC in $z$ direction. The absence of hinge states in the gap confirms the first-order topology of the system. (d) Bulk admittance dispersion at $R_c=0.2$ $\mathrm{mF}$ and $k_x=k_y=0$ with no WPs. (e) The admittance band structure under OBC along the $x$ direction and PBC in the $y$ and $z$ directions. There are two chiral modes near the locations of the original WPs. (f) The admittance dispersion under OBC in the $x$ and $y$ directions and PBC in the $z$ direction. No chiral hinge states emerge in the spectra. The common parameters are $C_1=-3.5$ $\mathrm{mF}$, $C_2= 2$ $\mathrm{mF}$,  $R_1=\sqrt{2}$ $\mathrm{mF}$ and $C_z=0.5$ $\mathrm{mF}$. } 
\label{gfig5}
\end{figure*}
	
We will now construct a TE circuit model that hosts not only various Weyl semimetal phases with edge and hinge states, but also exhibits first- and second-order chiral states when OBCs are applied to one and two dimensions, respectively. For this purpose, we incorporate finite non-reciprocal ($R_1$) and chiral ($R_c$) resistive couplings into the circuit. From Eq. \ref{WSMham}, we obtain the $k_z$-dependent admittance dispersion at ($k_x=k_y=0$) as 
\begin{widetext}
\begin{equation}
\mathcal{J}(\omega, k_z)=\pm \omega R_1 \pm \omega \sqrt{2((C_1+C_2)^2+2C_z^2+ R_c^2+ 4(C_1+C_2)C_z \cos k_z +(2 C_z^2-R_c^2) \cos (2 k_z))}.
\label{eq9}
\end{equation}
\end{widetext}
A finite $R_c$ and $R_1$ breaks the chiral, time reversal, $M_x$, and $M_z$ symmetries while preserving the $M_y$, $M_{xy}$ and inversion symmetries. The broken TRS symmetry allows for the emergence of Weyl points (WPs) in the admittance dispersion. When $R_1 \neq 0$, the system hosts a pair of WPs at $(0,0,k_z^{Weyl})$, where $k_z^{Weyl}$ is the resistive element-dependent position of the WPs on the $k_z$ axis and is given by
\begin{widetext}
\begin{eqnarray}
k_z=\pm \cos^{-1} \left( -\frac{(C_1+C_2)C_z}{2C_z^2-R_c^2} \pm \frac{1}{2} \frac{2 C_z^2 R_1^2 +R_c^2 (2 C_1+C_2)^2-8 C_z^2-R_1^2+4 R_c^2)}{(2C_z^2-R_c^2)^2}  \right). 
\label{eq10}
\end{eqnarray} 
\end{widetext}
To explain the role of the non-reciprocal resistive element $R_1$ on the behavior of the WP, we can further simplify Eq. \ref{eq10} by considering the case where $R_c=0$:
\begin{eqnarray}
k_z=\pm \cos^{-1} \left( -\frac{2(C_1+C_2)+\eta \sqrt{2}R_1}{4C_z}  \right), 
\label{eq11}
\end{eqnarray}
where $\eta$ can take the values of $\pm 1$. The two possible values for $\eta$ correspond to two different types of Weyl points with distinct Fermi arc behavior which we refer to as first-order and second-order Weyl points for reasons that will become apparent shortly. Depending on whether real solutions for $k_z$ exist  for only $\eta=1$, only $\eta=-1$, or both $\eta= \pm 1$,  we will obtain pairs of only first-order WPs, only second-order WPs, or both types of WPs, respectively.

We first consider the case where Eq. \eqref{eq11} has real solutions only for $\eta =+1$, which results in a pair of first-order WPs on the $k_z$ axis, as illustrated in the bulk (infinite) admittance dispersion of Fig. \ref{gfig5}a.  These WPs are called first-order WPs because they give rise to only first-order topological states, as can be seen by plotting the admittance spectra of the system in  a nanoplate geometry with finite width along the $x$-axis (see Fig. \ref{gfig5}b) and a nanowire geometry with finite widths along the $x$ and $y$ directions (see Fig. \ref{gfig5}c). It is evident that in the dispersion for the nanoplate geometry, the two WPs are connected by  edge states with nearly flat dispersion (see Fig. \ref{gfig5}b). However, for the nanowire geometry, the admittance bands become gapped (see Fig. \ref{gfig5}c), indicating the absence of second-order hinge states and the conventional or first-order nature of the WPs in the circuit lattice.

We now consider the influence of non-zero $R_c$ on the first-order topological states, the dispersion spectra of which are shown in the bottom row of Fig. \ref{gfig5} for the (infinite) bulk, nanoplate, and nanowire geometries. The bulk dispersion spectra become gapped and the system acquires mass-like terms for finite $R_c$. For the nanoplate geometry, two chiral modes appear in the admittance spectra of the circuit which cross the bandgap and meet at $k_x=0$ (see Fig. \ref{gfig5}e). However, for the nanowire geometry, the states remain gapped, and no mid-gap hinge states emerge  (Fig. \ref{gfig5}f). 
\begin{figure*}[ht!]
\centering
\includegraphics[width=0.85\textwidth]{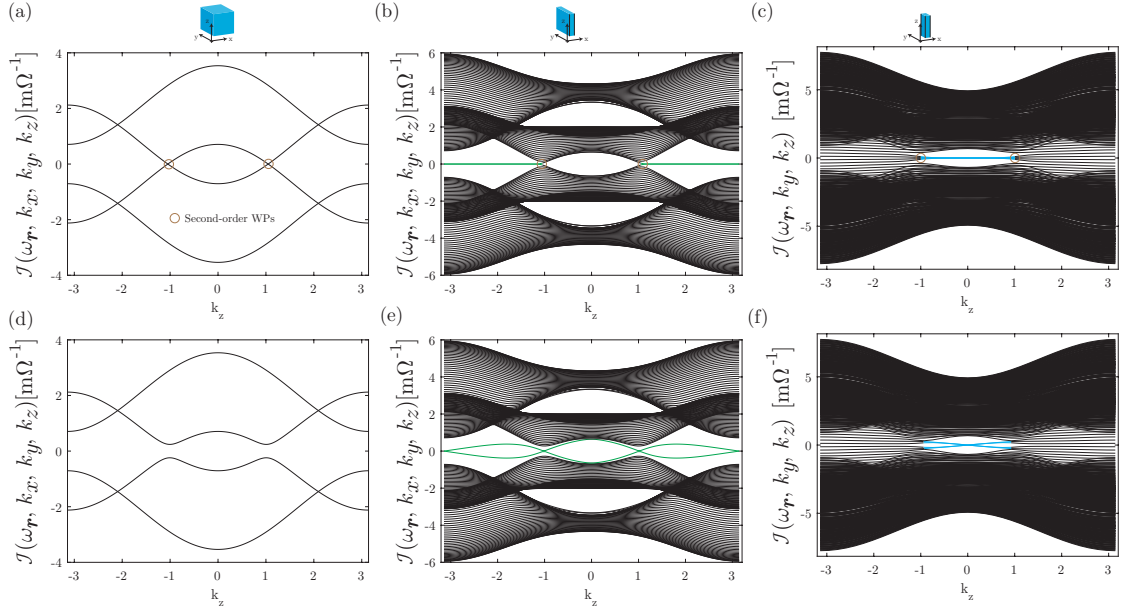}
\caption{ Evolution of the admittance band structure of the circuit model in Eq. \ref{WSMham} with non-zero resistive elements. (a) When $R_c=0$ and $k_x=k_y=0$, the low-admittance bands cross each other at two second-order WPs in the Brillouin zone at $\eta=-1$ only. (b) The admittance spectra under OBC in $x$ and PBC in $y$ and $z$ directions with $k_y=0$. The two WPs are connected by zero-admittance surface states. (c) The admittance dispersion under OBC in the $x$ and $y$ directions and PBC in the $z$ direction. The emergence of flat hinge states in the gap indicates the second-order topology in the system. (d) Bulk admittance dispersion of the circuit having broken chiral symmetry with $R_c=0.2$ $\mathrm{mF}$ and $k_x=k_y=0$, at which no WPs are found. (e) The admittance band structure under OBC in the $x$ and PBC in the $y$ and $z$ directions with $k_y=0$. There are two chiral modes near the original locations of the WPs. (f) The admittance dispersion under OBC in the $x$ and $y$ directions and PBC in the $z$ direction. Two chiral hinge states emerge in the spectra. The common parameters are $C_1=-1.5$ $\mathrm{mF}$, $C_2= 2$ $\mathrm{mF}$,  $R_1=\sqrt{2}$ $\mathrm{mF}$ and $C_z=0.5$ $\mathrm{mF}$. } 
\label{gfig6}
\end{figure*}

If Eq. \ref{eq11} has solutions only for $\eta=-1$, we obtain two second-order WPs in the bulk band (see Fig. \ref{gfig6}a). The second-order topology of the WPs can  be seen from the following: When OBCs are applied only along the $x$-direction, zero-admittance surface states that connect the two WPs emerge in the nanoplate geometry. When a finite width is further introduced into the $y$ direction to form  a nanoribbon geometry with finite widths along $x$ and $y$ directions (Fig. \ref{gfig6}b) , the first-order edge states become quantized, and second-order hinge states that cross the bulk bandgap emerge (Fig. \ref{gfig6}c, respectively) In the presence of a finite $R_c$, the bulk band will have a finite admittance gap (see Fig. \ref{gfig6}d). For the case of finite $R_c$,  chiral surface and hinge states emerge in the midgap  of the admittance spectra for the nanoplate and nanoribbon geometries, as shown in Fig. \ref{gfig6}b and \ref{gfig6}c, respectively.  We dub these chiral-type hinge states as a second-order chiral hinge states. 

In summary, we have realized both topological first- and second-order chiral states and Weyl semimetallic phases by varying electrical resistive parameters in a TE circuit without the requirement of any external magnetic fields. In higher-order topologically non-trivial systems (which include topological insulators and Weyl and Dirac semimetals), the bulk of the system, as well as surfaces of the 3D model are all insulating but the hinges of the model are conducting, i.e., they host distinct hinge states.  The hinge states in particular are expected to be robust against perturbations because the direction of their propagation is locked to their pseudospin. They are also applicable to the study of Majorana fermions, which are actively being investigated for applications in fault-tolerant quantum computing \cite{you2019higher}. This robust unidirectional property, in which current flow is allowed only one direction along a hinge, implies that a chiral hinge current excited at one hinge in a cuboid circuit cannot flow into another hinge situated diagonally opposite from the hinge being excited \cite{ni2020robust,okugawa2019second}. This property can therefore be exploited for robust topological signal multiplexing by utilizing the multiple discrete degrees of freedom in the system \cite{mei2019robust}. Finally, HOWSM states with hinge states open the possibility for robust dissipationless interconnects \cite{hofmann2019chiral} and analogues of truly 1D superconducting nanowires. \cite{langbehn2017reflection}.

\section{Conclusion}
In this paper, we proposed a tunable scheme to realize various higher-order topological gapless and chiral phases in a TE network consisting of basic electrical components such as resistors, inductors and capacitors. We first constructed the circuit model for a two-dimensional (2D) higher-order topological insulators using these basic components. We then extended the original 2D TE circuit in the vertical z-direction by stacking copies of the 2D circuits one on top of another. By coupling the nodes in adjacent layers diagonally using a common stacking capacitor $C_z$, we can modulate the intra- and intercell coupling of the effective 2D Laplacian to realize a richer set of topological properties associated with the gapless states in three dimensions (3D). For instance, we obtain a flat band with higher-order hinge states that connect two gapless nodes together. The gapless nodes exhibit Dirac or Weyl semimetalic characteristics depending on the circuit symmetries. Interestingly, the chiral symmetry of the hinge states can be broken by adding resistive couplings between equivalent nodes on adjacent layers. In this case, the admittance spectrum becomes gapped, but the two hinge states survive and propagate with positive and negative group velocities in the z-direction. Furthermore, by incorporating tilting capacitances, both chiral modes in a given valley can be made to propagate in same direction but opposite to that of the corresponding modes in the other valley. The flat-band edge and hinge states in these 3D layered TE circuits may find applications in sensors with high sensitivity and ultra-low dissipation owing to their tunability and chirality.

\section{Appendix}

\subsection{ Negative resistance converter with current inversion (INIC) and dynamic stability of Op-amps}
\begin{figure}[ht!]
\centering
\includegraphics[width=0.46\textwidth]{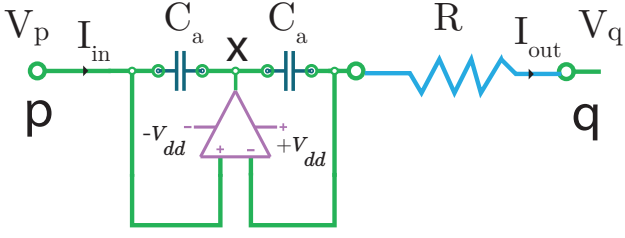}
\caption{ Illustration of a simple negative resistance converter with current inversion (INIC). } 
\label{gFigs1}
\end{figure}
To induce a negative imaginary onsite potential (i.e., gain term) in the TE circuit array, we use the unity gain operational amplifier (op-amp) circuit shown in Fig. \ref{gFigs1} to provide an additional $\pi$ phase modulation with respect to the original resistance value. The circuit comprises two feedback capacitors with the same capacitance $C_a$, an operational amplifier, and a resistance $R$. The relation between the input current and the voltage for the forward (node $p$ to $q$) and backward (node $q$ to $p$) directions can be expressed in the matrix form 	
\begin{equation}
	\begin{pmatrix} I_{pq} \\ I_{qp} \end{pmatrix} = \frac{1}{R} \begin{pmatrix} -1 & 1 \\ -1 & 1 \end{pmatrix} \begin{pmatrix} V_p \\ V_q \end{pmatrix},
	\label{seq01} 
\end{equation}
where $I_{ij}$ and $V_i$ denote the current passing from the $i$th to the $j$th node and the voltage at the $i$th node respectively. From Eq. \ref{seq01} we can easily obtain
\begin{equation}
I_{pq}=-I_{qp}=- \frac{V_p-V_q}{R}.
\end{equation}
Therefore, for the coupling from node $p$ to $q$, the resistance will acquire a phase of $\pi$ relative to the coupling from node $q$ to $p$, and behave as a negative resistor with value of $-R$.  If we replace resistance by inductor or capacitor or combination of resistance and capacitor, Fig. \ref{seq01} translate to a general schematic of a negative impedance converter with current inversion (INIC).
 
\subsection{Determination of winding number } 
As explained in the main text, the circuit Laplacian described in Eq. \ref{WSMham} at resonant frequency but without tilting capacitance (i.e, $C_t=0$) respects both chiral and mirror rotational symmetry around the $xy$ plane. The Laplacian can be transformed into a block diagonal form along a high symmetry line  (i.e., $k_x=k_y=k$) after a simple unitary transformation
\begin{eqnarray}
\chi ^{-1} Y (\omega_r,k,k_z) \chi = \begin{bmatrix}
Y_1 (\omega_r,k,k_z) & 0\\
0 & Y_2 (\omega_r,k,k_z)
\end{bmatrix},
\label{eqs2}
\end{eqnarray}
where $Y_1 (\omega_r,k,k_z)$ and $ Y_2 (\omega_r,k,k_z)$ are given by 
\begin{eqnarray}
\begin{aligned}
Y_{1}(\omega_r,k,k_z)= &(C_1+2C_z \cos k_z+C_2 \cos k) \sigma_x  \\ &+  C_2 \sin k \sigma_y,
\label{eqs3}
\end{aligned}
\end{eqnarray}
and 
\begin{eqnarray}
\begin{aligned}
Y_{2}(\omega_r,k,k_z)= &(C_1+2C_z \cos k_z+C_2 \cos k) \sigma_x \\ & -  C_2 \sin k \sigma_y,
\label{eqs4}
\end{aligned}
\end{eqnarray}
with $\chi$ is the unitary transformation matrix. If we consider $k_z$ as a tuning parameter, the winding number of  $Y_1 (\omega_r,k,k_z)$ can be obtained explicitly as 
\begin{align}
\mathcal{W}_1 & = -\frac{i}{2 \pi} \oint \psi^\dagger \partial_k \psi\ \mathrm{d}k  \nonumber \\
  &= \frac{1}{2\pi} \int_{k=0}^{k=2\pi}  \mathrm{d}k\ \left[ \tan^{-1} \left(\frac{\sin k}{\frac{C_1+2C_z \cos k_z}{C_2}+\cos k} \right) \right] \nonumber \\
  &= \theta (C_2-(C_1+2C_z \cos k_z)),
\label{eqs5}
\end{align} 
where $ \theta$ is the Heaviside step function and $\psi$ is the eigenstates of $Y_{1}(\omega_r,k,k_z)$. Similarly, the winding number of $Y_{2}(\omega_r,k,k_z)$ can be calculated as 
\begin{equation}
\mathcal{W}_2 =-\theta (C_2-(C_1+2C_z \cos k_z)).
\label{eqs6}
\end{equation}
Therefore, the resultant winding number of the $4 \times 4$ matrix in Eq. \ref{WSMham} of the main text is given by 
\begin{equation}
\mathcal{W}= \mathcal{W}_1-\mathcal{W}_2=
\begin{cases}
  2, & \text{if }
       \!\begin{aligned}[t]
       \bigg|\frac{C_1+2C_z \cos k_z}{C_2}\bigg| \leq 1 \\
            \end{aligned}
\\
  0, & \text{otherwise}
\end{cases},
\label{eqs7}
\end{equation}
as shown in Eq. \ref{eq6} of the main text.

\subsection{Distinguishing first order and second order gapless states via impedance spectra }
\begin{figure*}[ht!]
  \centering
    \includegraphics[width=0.75\textwidth]{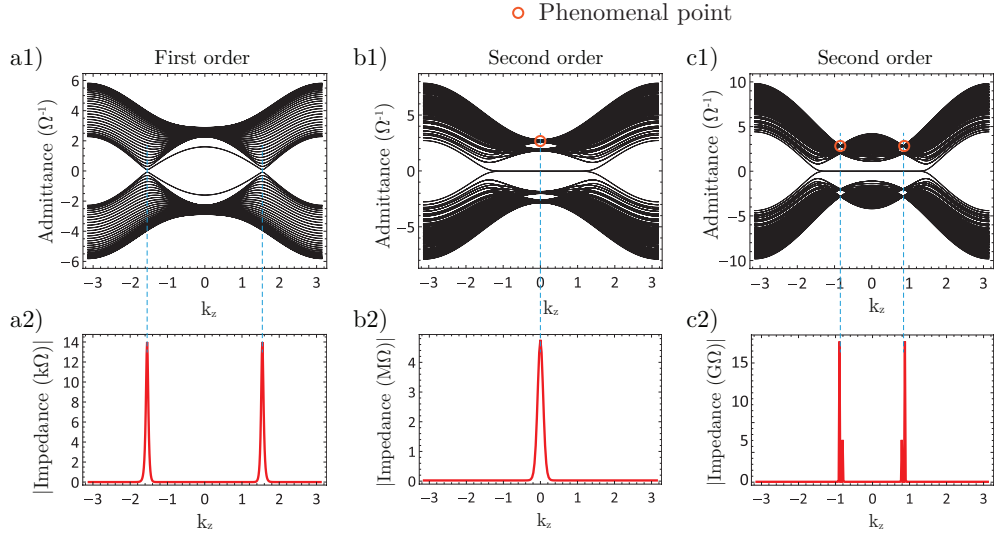}
  \caption{Admittance band structures and impedance spectra as a function of $k_z$. (a1, a2) when considering nanoribbon geometry consisting of 30-unit cells in the finite $x$-direction and PBC in the infinite $y$ and $z$ directions to characterize first order Weyl points, with $k_y=0$ and $C_z=0.95$ $\mu F$. (b1, b2, c1, c2) The admittance spectrum and impedance responses of the nanoplate geometry, with PBC along the $z$ direction and OBC in both the $x$ and $y$ directions to characterize second-order Weyl points. Phenomenal points are marked in orange circles for b1 and c1. Common parameters used are $C_1=-2$ $\mu F$, $C_2=2$ $\mu F$, however $C_z=0.95$ $\mu F$ for b1 and b2 and $C_z=1.5$ $\mu F$ for c1 and c2. }
  \label{gapp3}
\end{figure*}

High impedance peaks occur at the points where the gap closes, which correspond to the topological phase transition points. As implied by Eq. \eqref{eq7} of our manuscript, nearly zero eigenvalues result in a large impedance readout, allowing for the identification of topological zero modes. In the circuit model, the phase order can be determined based on the dissimilar eigenspectra in the impedance responses between first-order and second-order phases for a given parameter set. For example, the band touching points in the first-order semimetallic phase (Fig. \ref{gapp3}a1) occur only at $k_z = \pm \cos^ {-1}\left( -\frac{C_1+C_2}{2C_z} \right)$ . In contrast, the band closing point in the second-order phase occurs at $k_x=k_y=k_z=0$ if $|C_1+C_2 | < | 2 C_z |$. This means that one should observe a single impedance peak at $k_z=0$ in the second-order phase if there is no distinguishable phenomenal points or two degenerate phenomenal points (see Fig.  \ref{gapp3}b1) in the band structure whereas there are two impedance peaks at $k_z = \pm \cos^{-1}\left( - \frac{C_1+C_2}{2C_z} \right)$  in the first-order phase.

As for the second-order semimetallic phase where two distinct phenomenal points occur in the admittance band diagram corresponding to some other range of parameter values (see Fig.  \ref{gapp3} c1), the impedance peak at $k_z=0$ is now split into two sharp impedance peaks coinciding with the phenomenal points rather than band touching points (see Fig.  \ref{gapp3} c2). In Fig.  \ref{gapp3}, we present these three cases in which the impedance peaks provide signatures not only to determine whether the system is in a topologically trivial or non-trivial phase, but also to distinguish the different orders of the semimetallic phases.

As can be seen in Fig.  \ref{gapp3} a1 and a2, the peak impedance response of the circuit corresponds to the two Dirac points for the first-order topology under nanoribbon geometry. However, for the second-order topology, the peak impedance characteristics correspond to the phenomenal points. In the corresponding nanoplate geometry (Fig.  \ref{gapp3} b1 and b2), there is a single phenomenal point which translates to a single impedance peak at $k_z=0$. However, for the nanoplate geometry, there are two phenomenal points in the second-order topology, which translate into two impedance peaks (see Fig.  \ref{gapp3} c1 and c2).  Furthermore, the magnitude of the impedance peaks for the second-order semimetallic phases tend to have a much larger magnitude compared to the first-order counterpart. This is due to the presence of the almost zero energy line connecting two gapless points in the second-order topological regime (see Figs. b1 and c1 of Fig.  \ref{gapp3}). On the contrary, in the first-order semimetallic phase (Fig.  \ref{gapp3} a1), there is no such zero-admittance line between two gapless points, and hence we obtain comparatively low peak impedance values. Based on the above observations, it is thus possible to distinguish the different topological orders of the semimetallic phase via their impedance readout characteristics.

\subsection{Topological origin of higher order gapless states}
In Figs. \ref{gfig5} and \ref{gfig6} of the main manuscript, we discussed the first and second order weyl semimetalic phases. Interestingly, the  surface and higher order topological states can be characterized via the   Chern number and quantized quadrupole moment index, respectively. To elaborate, when time reversal symmetry is broken by finite $R_1$ but inversion symmetry is preserved, two first-order Weyl points (WPs) appear in the energy band dispersion (see Fig. \ref{gfig5} of the main manuscript) at a certain parameter region and two second-order Weyl points emerge (see Fig. \ref{gfig6}a of the main manuscript) at other parameter regions. Interestingly, the two second-order WPs appear at the boundaries of the $k_z$ regions within which hinge states (see Fig. \ref{gfig6}a). The second-order WPs are characterized via the quantized quadrupole moment index \cite{benalcazar2017electric,ghorashi2020higher}, which is defined as
\begin{equation}
q_{xy} (k_z )=(p_x  (k_z )+ p_y  (k_z )- Q_c  (k_z )  mod 1)
\label{appd1}
\end{equation}
where $p_x  (k_z )$ and $p_y  (k_z )$ are the $x$ and $y$ components of the polarization on the surfaces normal to the $y$ and $x$ directions, respectively. $Q_c (k_z)= \int_{v}(-\vec{\Delta}. \vec{\rho})dv$, is the corner charge \cite{benalcazar2017electric}, where $\rho$ denotes the volume charge density. In the above equation, we define
\begin{equation}
p_x  (k_z )  = \frac{1}{(2\pi)^2} \int_{\mathrm{BZ}} \mathrm{Tr}[A_{x,k}]\ \mathrm{d}^2 k, 
\label{appd2}
\end{equation}
and 
\begin{equation}
p_y  (k_z )  = \frac{1}{(2\pi)^2} \int_{\mathrm{BZ}} \mathrm{Tr}[A_{y,k}]\ \mathrm{d}^2 k, 
\label{appd3}
\end{equation}
where $[A_{i,\mathbf{k}}]^{mn}=-i \langle u_{\mathbf{k}}^m | \partial_{k_i} | u_{\mathbf{k}}^n \rangle $ is the Berry connection and $u_{\mathbf{k}}^m$ is the $m$th eigenvector.
Explicitly, $q_{xy} (k_z)$ can be expressed as 
\begin{equation}
q_{xy} (k_z ) =- \frac{1}{2}, \quad  k_z^{w_l^{2nd}}\leq k_z  \leq k_z^{w_r^{2nd}}
\label{appd4}
\end{equation}
where  $k_z^{w_l^{2nd}}$ and $k_z^{w_r^{2nd}}$ are the two solutions of $k_z$ that are given in Eq. \ref{eq11} with $\eta=-1$ and $R_c=0$ and can be expressed explicitly as 
\begin{equation}
k_z^{w_l^{2nd}}= -\cos^{-1} \left( -\frac{2(C_1+C_2)-\sqrt{2}R_1}{4C_z} \right),
\label{appd5}
\end{equation}
and 
\begin{equation}
k_z^{w_r^{2nd}}= \cos^{-1} \left( -\frac{2(C_1+C_2)-\sqrt{2}R_1}{4C_z} \right).
\label{appd6}
\end{equation}
In contrast, first-order WPs appear where the Fermi arc-like surface states exist and can be characterized by the Chern number \cite{benalcazar2017electric}, which is defined as 
\begin{equation}
C=\frac{1}{2\pi} \int_{BZ} \mathrm{d}^2k\  \mathrm{Tr}[\partial_{k_x} A_{y,k}-\partial_{k_y} A_{x,k}]. \label{appd7}
\end{equation} 
Hence, explicitly, the Chern number can be calculated as 
\begin{equation}
C =1, \quad  k_z^{w_l^{1st}}\leq k_z  \leq k_z^{w_r^{1st}},
\label{appd8}
\end{equation}
where,  $k_z^{w_l^{1st}}$ and $k_z^{w_r^{1st}}$ are the two solutions of $k_z$ that are given by the Eq. \ref{eq11} with $\eta=+1$ , $R_c=0$ and can be expressed explicitly as 
\begin{equation}
k_z^{w_l^{1st}}= -\cos^{-1} \left( -\frac{2(C_1+C_2)+\sqrt{2}R_1}{4C_z} \right),
\label{appd9}
\end{equation}
and 
\begin{equation}
k_z^{w_r^{1st}}= \cos^{-1} \left( -\frac{2(C_1+C_2)+\sqrt{2}R_1}{4C_z} \right).
\label{appd10}
\end{equation}
In other words, first-order Weyl points appear on the phase boundary between the trivial and $C=1$ phases while second-order WPs appear on the phase boundary between the $C=1$ and $q_{xy} (k_z )=-\frac{1}{2}$ phases. A schematic phase diagram is shown in figure below
\begin{figure*}[ht!]
  \centering
    \includegraphics[width=0.75\textwidth]{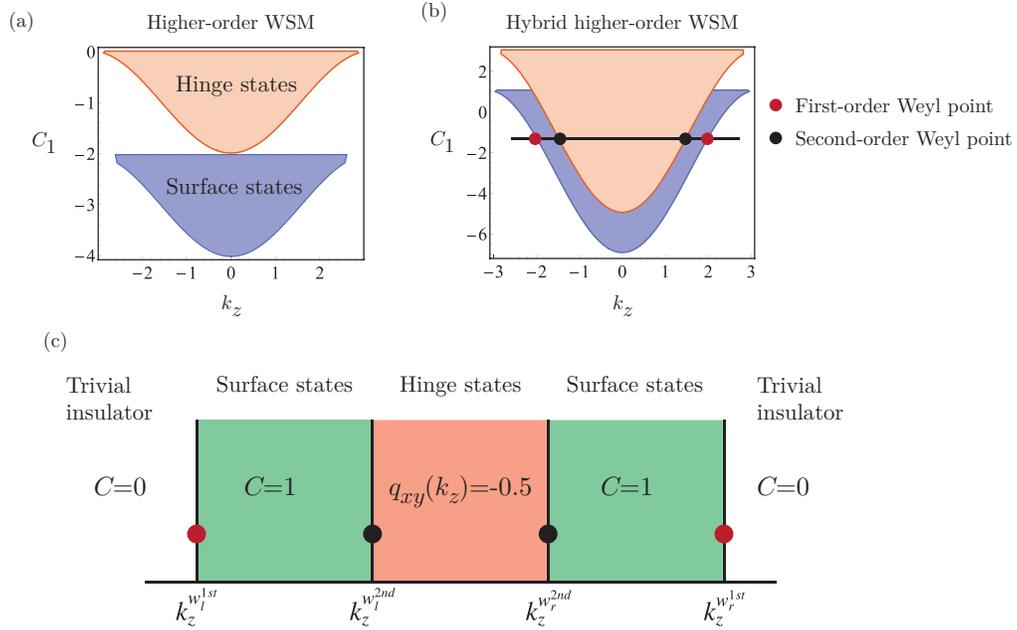}
  \caption{Phase diagram for various Weyl semimetallic phase. a. Phase diagram in $k_z-C_1$ plane with $C_2=2$ mF, $C_z=0.5$ mF, $R_1=\sqrt{2}$ $k \Omega$ and $R_c=0$.  In different parameter regions, we get either hinge or surface states indicating the presence of second-order or first-order WPs, respectively. b. Phase diagram in $k_z-C_1$ plane with $C_2=2$ mF, $C_z=2$ mF, $R_1=\sqrt{2}$  $k \Omega$ and $R_c=0$.  In some parameter ranges, we obtain both hinge and surface states simultaneously, indicating a hybrid higher-order topological system. c. Simplified phase diagram with respect to $k_z$ calculated using Eq. \ref{eq10} of the main manuscript. }
  \label{gapp4}
\end{figure*}  
In Fig. \ref{gapp4}, we plot the phase diagram for our circuit model. Depending on the solutions of Eq. \ref{eq11}, we obtain first- or second- order Weyl points or Weyl points of both orders.  The light orange and blue shaded areas in Fig. \ref{gapp4}(a) denote the solution of Eq. \ref{eq11} with   $\eta=-1$  and $\eta=+1$, which indicate the presence of hinge and surface states, respectively. Since there is no overlap between two regions, there are either only surface states or only hinge states depending on the parameter choice.  On the other hand, at some parameter ranges, the overlap between the light orange and blue shaded areas denotes the simultaneous presence of surface and hinge states indicating a hybrid-order Weyl semimetalic phase. A schematic phase diagram on the $k_z$ axis is shown in panel c.  Do note that in the presence of non-zero $R_c$, the solutions for $k_z^{w_{l/r}^{2nd}}$ and  $k_z^{w_{l/r}^{1st}}$ can be found using Eq. \ref{eq10}. 

\subsection*{Data availability}
The data that support the plots and other
findings of this study are available from the corresponding author upon reasonable request.

\subsection*{Code availability}
The computer codes used in the current study are accessible from the corresponding author upon reasonable request.
\subsubsection*{Acknowledgements}
This work is supported by the Ministry of Education (MOE) of Singapore Tier-II Grant MOE-T2EP50121-0014 (NUS Grant No. A-8000086-01-00), and MOE Tier-I FRC Grant (NUS Grant No. A-8000195-01-00).

\subsection*{Author contributions}
S.M.R-U-I, Z.B.S, H.S and M.B.A.J initiated the primary idea. S.M.R-U-I and Z.B.S  contributed to the formulation of the analytical model, code development, data analysis, and to the writing of the manuscript under the kind supervision of  M.B.A.J.

\subsection*{Additional information}
\textbf{Competing interests:}  The authors declare no competing interests.


%

\end{document}